# Structured Descriptions of Roles, Activities, and Procedures in the Roman Constitution

Yoonmi Chu and Robert B. Allen

Yonsei University, Seoul, Korea
yoonmichu@gmail.com and rba@boballen.info

**Abstract.** A highly structured description of entities and events in histories can support flexible exploration of those histories by users and, ultimately, support richly-linked full-text digital libraries. Here, we apply the Basic Formal Ontology (BFO) to structure a passage about the Roman Constitution from Gibbon's *Decline and Fall of the Roman Empire.* Specifically, we consider the specification of Roles such as Consul, Activities associated with those Roles, and Procedures for accomplishing those Activities.

**Keywords:** BFO, Community Models, Digital Humanities, Granularity, Ontologies, Procedures, Roles

## 1 Introduction

### 1.1 Full-text Digital Libraries and Community Models

Full-text digital libraries could incorporate rich descriptions which allow flexible linking and interaction with the content. [2] showed how a structured passage of Gibbon's well-known Decline and Fall of the Roman Empire could be used in a novel content browser. A formal description of Roman culture and legal systems would be useful because the organization of the government of the Empire and the rights of its citizens are keys for Gibbon's analysis. Not only has no one proposed a systematically structured description for the Roman Constitution; there are many open issues about how such a framework should be constructed. Here we focus on Roles, Activities, and Procedures because they are central to such a description.

This work also contributes to broader efforts to describe cultural and historical information. Our approach goes beyond using metadata to describe historical documents and artifacts; we describe the dynamic and systematic structures of entities appearing in history as reflected in various writings and other materials. Some of our earlier work examined the description of text from digitized historical newspapers. The newspapers provided relatively succinct descriptions of a broad range of activities of communities in small towns. While some of the events in the newspapers were disjoint we believe they could be unified through modeling the communities on which the newspapers report. Potentially, such community models would allow continuity and linking across events as the basis for a full-text digital library and could also link in a wide range of non-textual materials.

### 1.2 Upper Ontologies

We apply ontologies with rich semantic structure to model complex entities in historical content. Upper ontologies, particularly the BFO (Basic Formal Ontology)[1], provide a typology of entities which is domain neutral. The BFO is especially well developed and is widely used in biology but it has not previously been applied to cultural-heritage materials. BFO is considered as "realist" ontology. DOLCE [9, 10] is another widely used upper ontology which focuses on linguistic and cognitive entities. While most work in digital humanities and digital history has used DOLCE (e.g., [7, 13]), we decided to explore the BFO because the entities we wished to model were explicit social structures such as those described in a constitution or used in a fact-based news article. In some cases, social structures can be highly nuanced and subjective; indeed, modeling them can become highly contentious.

[1] http://www.ifomis.org/bfo

BFO distinguishes between continuants (or endudrants) which persist through time and occurrents (or perdurants) that occur in time and unfold across time, e.g., processes, events, activities and changes [6]. The BFO creates separate ontologies for continuants and occurrents called SNAP and SPAN respectively. A SNAP ontology represents a snapshot of the state of reality which is composed of continuants. By comparison, a SPAN ontology applies to the reality constituted by processual entities that unfold across spatiotemporal and temporal regions. In other words, SPAN ontologies are four dimensional (4D) [6]. Additionally, BFO defines SNAP-SPAN trans-ontology relationships to coordinate these sub-ontologies in a coherent framework. Historical information should be well described because the SNAP entities can depict states of continuants and can participate in a SNAP-SPAN trans-ontology as bearers of occurrents.

## 2 Roles, Activities, Processes, and Procedures

The BFO includes Roles in the "specifically dependent continuant" class. A BFO Role is a realizable entity that an independent continuant can take on but does not reflect the physical structure of that independent continuant [3]. This means that Roles are optional and their manifestation is a reflection of surrounding circumstances [8]. Separating Roles and the bearer of Roles allows for an individual to hold multiple Roles. The performance of a specific Role is determined by conditions and situations of its bearer, thus we should be able to model the context. [8] proposed a Role representation model to deal with OWL axioms and SWRL rules.

Roles in the BFO have usually been applied to biological processes such as the Role of DNA in reproduction. They have not been applied to the humanities[2]. Social roles imply privileges (or rights as a broader sense) and obligations (or norms or responsibilities as a broader sense) depending on how the Role is specified. For example, a person has the right to vote in an election of the committee president as a member of a committee and at the same time he/she is expected to pay taxes as a citizen.

In BFO Processes are processual entities in a SPAN ontology that are occurrents or happenings located at temporal or spatiotemporal regions [6]. They involve SNAP entities as their participants and are dependent on their participants. Simple processes are continuous ongoing activity such as "running". Processes in SPAN can be both "dissective" (composed of other processes) and "cumulative" [6].

[2] DOLCE [10] defines social objects that are further divided into Agentive and Non-agentive but it does not address the complexity of the social objects and interaction between them.

We define Procedures, separating from Processes, as an "inherent" attribute of social objects such as operating rules in an organization. According to [4], Procedures are similar to closed processes that consist of a definite sequence of actions or activities leading to a specific result. We can describe Procedures with semantic structures defined in BFO. We believe that Procedures are continuants, since they are complex attributes inherent in their bearer and are just specifications of some things that could happen. A Procedure which is composed of Activities is defined as a subclass of *realizable entity* in contrast with Processes which are occurrents[3].

## 3 Application to the Roman Constitution

Gibbon's famous historical analysis considered the changes in Roman society during the Roman Empire [5]. Gibbon's text includes a large number of social concepts as well as their relationships and interaction. To explore the issues for modeling complex entities and processes we chose a sample passage dealing with Roman Constitution[4]:

> The consuls had succeeded to the kings of Rome, and represented the dignity of the state. They superintended the ceremonies of religion, levied and commanded the legions, and presided in the assemblies both of the senate and people. … but whenever the senate empowered the first magistrate to consult the safety of the commonwealth, he was raised by that decree above the laws, and exercised, in the defence of liberty, a temporary despotism.[5]

In Figure 1, we show a partial ontology to represent the concepts of Roles, Rights and Activities of the Senate and the Consuls that are the highest magistrate in the Constitution of Roman Empire. A person who is the Senate has *senate role* and a person who is the Consuls has *first magistrate role* belonged to *consul role*. In this representation, each Role has some Rights, which consist of Activities. If a Role has a Right and the Right that the Role owns has an Activity, then the Role is permitted to perform the Activity. This can be expressed as a SWRL rule and permitted activities of the Role can be inferred it:

**rule_1:** *role(?x) ^ rights(?y) ^ activity(?z) ^ has right(?x, ?y) ^ has activity(?y, ?z) -> permitted activity(?x, ?z)*[6]

Separating Rights and Activities enables us to dynamically add or delete permitted activities of the Role and also to compose multiple Roles and Rights. Thus, we can describe the variability of Rights that a Role has according to time or context [8].

---

[3] In forthcoming work we also define Workflows. Procedures and Workflows may be related to the Information Artifacts ontology (IAO) and the Software Ontology (SWO). The authority to exercise those, as well as other aspects of the Constitution depends on the interpretation of the basis of social authority which goes back to the Smith/Searle debate.

[4] The Roman Constitution was not a single document but a set of laws. In addition, the Constitution evolved over time – especially in the transition from Republic to Empire.

[5] From Gibbon Chapter III: The Constitution In The Age Of The Antonines.—Part I.

[6] This rule can also be expressed with property chains in OWL2.

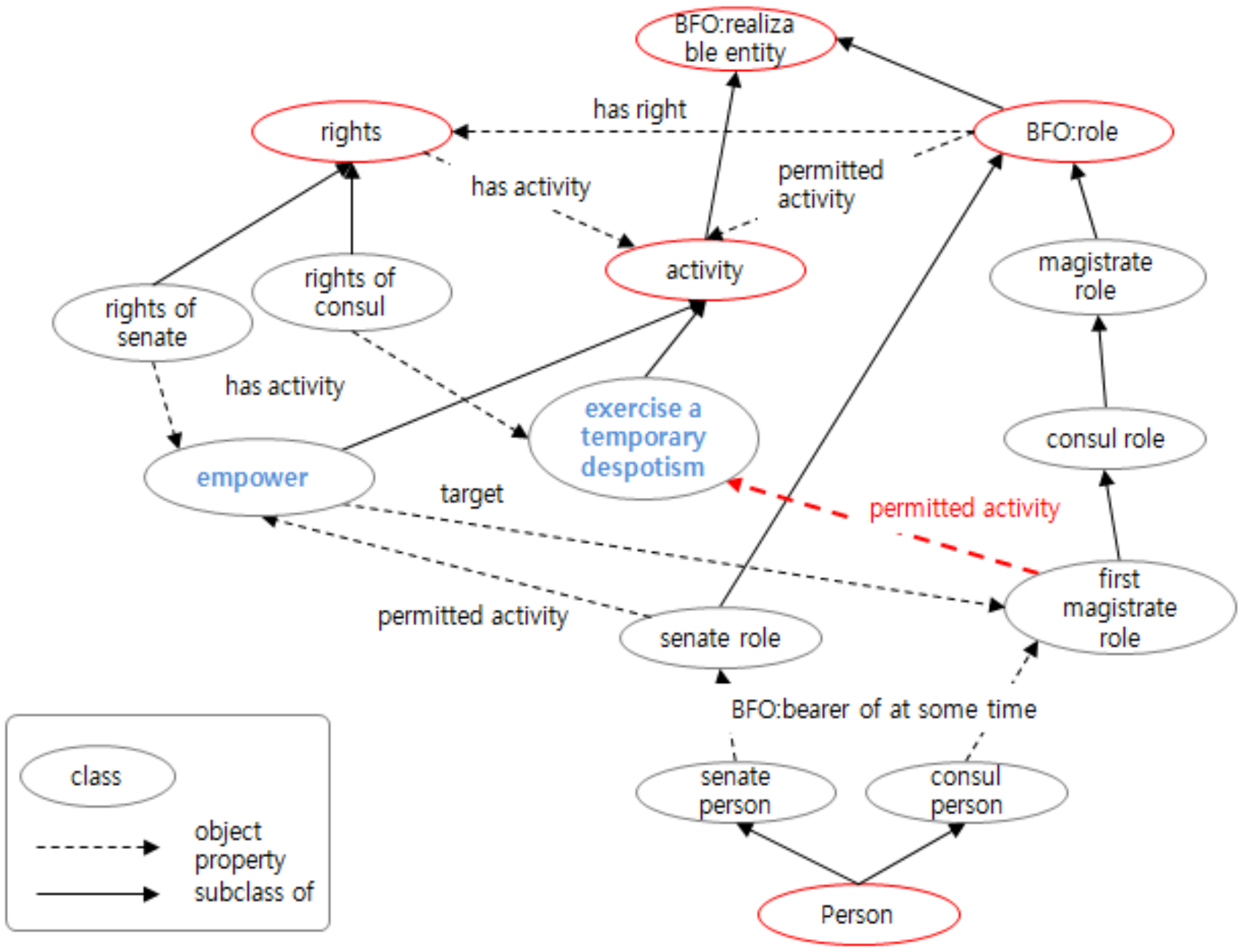

**Fig. 1. Representation of the Activities in wartime as specified by Gibbon.**

The final sentence of the sample passage can be regarded as a Procedure which consists of some Activities with sequences to be executed. According to the sentence, the Senate has the right to empower the First Magistrate in wartime. This right could be activated only when the safety of the Roman Commonwealth was threatened.

Figure 2 shows the structure of Processes, Procedures and Activities that compose it. The *wartime* Procedure contains *empower* Activity and *exercise a temporary despotism* Activity, while the former is a prerequisite of *exercise a temporary despotism* Activity. The *prerequisite of* property describes the sequences of these activities. The *wartime* Process, which is a historical event having spatiotemporal regions[7], is associated with the *wartime procedure* with *realized in* property defined in the SNAP-SPAN trans-ontology. This ontological model enables us to describe historical information more semantically and systematically.

## 4 Discussion and Conclusion

We have examined applying the BFO to the description of Roles, Activities, Procedures and Processes such as required for the description of the Roman Constitution and, more broadly, for governmental structures of communities. With BFO, we can describe complex entities in a 4D perspective. We defined Procedures which consist of activities, distinguishing Procedures from Processes. They are inherent in objects which are its bearers as a complex type of attributes, and can be realized by Processes.

[7] To focus on the relation of Processes and Procedures having Activities, we simplified the model by excluding spatiotemporal or temporal regions.

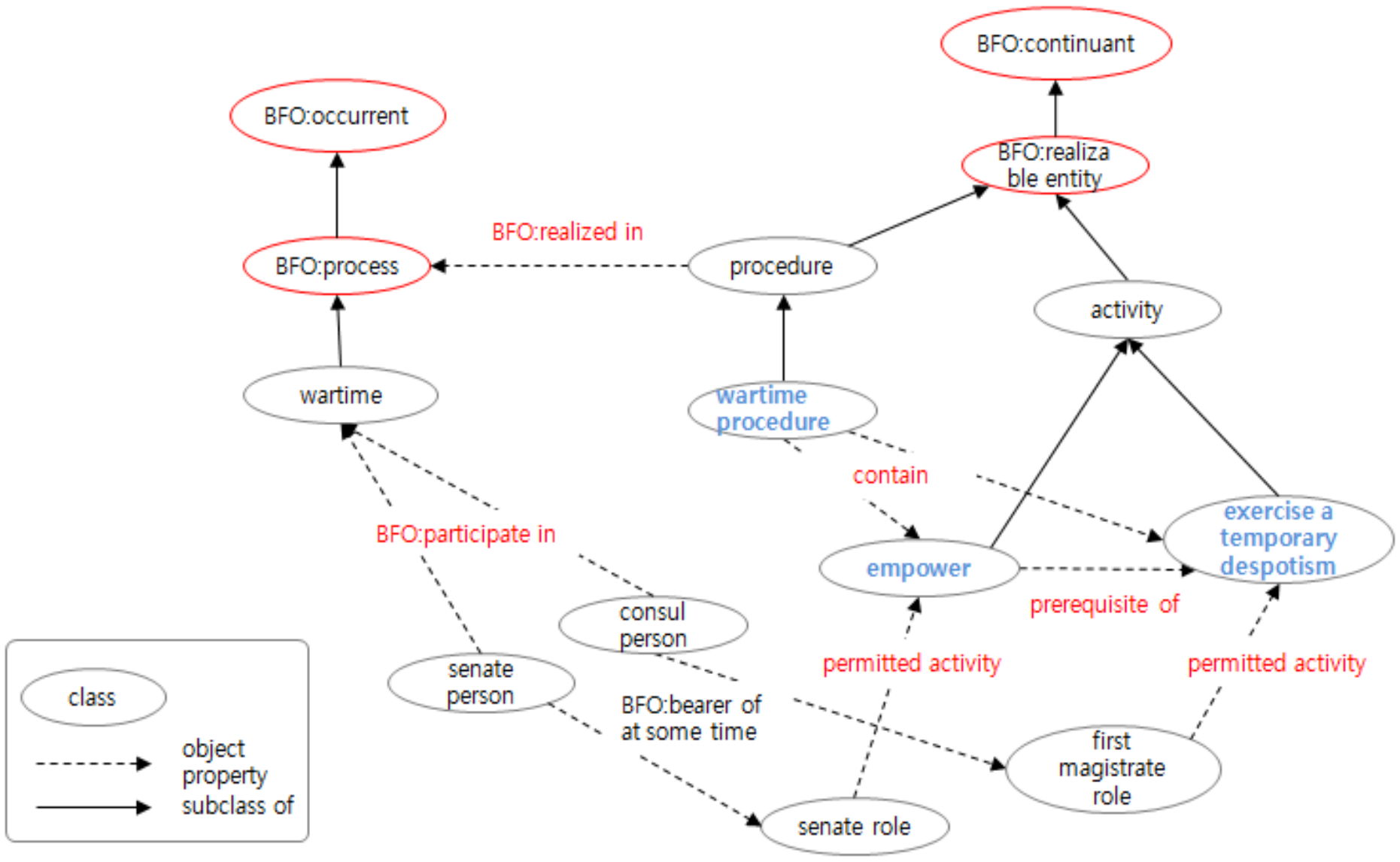

**Fig. 2. Example of the relation between Processes and Procedures.**

We did not deal with complex workflows in our example; workflows offer several puzzles in terms of ontological modeling. Even the Wartime Procedure described above has an aspect of workflow since it is granted conditional on whether the Empire is at war. Workflows have same features of programming languages such as conditionals and looping. While ontologies can represent individual states, they do not typically represent full state machines. Rather, state machines are most often associated with programming languages. Some of these issues are covered by a proposed BFO structure known as a Process Profile[8]. However, there is relatively little discussion of it and no formal structures have proposed for it. We suggest that its implementation requires concepts such as looping which are integral to programming languages and business process engineering. While importance of programming language features for ontologies is recognized in SPIN[9] and OWL-S,[10] these are for restricted applications and we believe that a broader integration is needed (see [2, 12]).

The AI and Law community has considered ontological models for legal concepts which combine lexical and conceptual meaning [11, 14]. [1] showed the possibility that a legal concept like 'public function' could be modeled in ontologies using FrameNet knowledge. It also showed interconnecting 'LegalRole' and 'Action' through the case study of the 'obligation' concept. We believe there is the opportunity for more work on building ontological models of complex concepts like those we study here.

---

[8] It describes repeating patterns such as the beating of a heart in BFO2; http://purl.obolibrary.org/obo/bfo/2012-07-20/bfo.owl

[9] http://www.w3.org/Submission/spin-overview/

[10] http://www.w3.org/Submission/OWL-S/

Although this study has limitations, and there are complex questions such as the difficulty of tracking the changes to the Roman Constitution as it evolved through time, it suggests the potential for structured descriptions which can be useful for representing, linking and discovering of historical knowledge. By exploring some of the issues for applying Roles, Processes and Procedures, it provides a modeling framework to describe social objects as complex entities in the construction of Community Models and broader Societal Models.

## 5 References

1. Agnoloni, T., Barrera, M. F., Sagri, M. T., Tiscorni, D., Venturi, G.: When a FrameNet-style Knowledge Description Meets an Ontological Characterization of Fundamental Legal Concepts. In: AI Approaches to the Complexity of Legal Systems. Complex Systems, the Semantic Web, Ontologies, Argumentation, and Dialogue. pp. 93-112. Springer, Berlin Heidelberg (2010)
2. Allen, R.B., Chu, Y.M..: Towards a Full-Text Historical Digital Library, In: ICADL, LNCS 8839, pp. 218-226 (2014)
3. Arp, R., Smith, B.: Function, Role, and Disposition in Basic Formal Ontology. Nature Proceedings, 1941(1), pp. 1-4 (2008)
4. Galton, A.: Experience and History: Processes and their Relation to Events. Journal of Logic and Computation. (2008)
5. Gibbon, E.: The History of the Decline And Fall of the Roman Empire, (1845) http://www.gutenberg.org/files/731/731-h/731-h.htm
6. Grenon, P., Smith, B.: SNAP and SPAN: Towards Dynamic Spatial Ontology. Spatial Cognition and Computation 4(1), pp. 69-104 (2004)
7. Grossner, K. E.: Representing Historical Knowledge in Geographic Information Systems, Doctoral dissertation, University of California, Santa Barbara (2010)
8. Kozaki, K., Sunagawa, E., Kitamura, Y., Mizoguchi, R.: Role Representation Model Using OWL and SWRL. In: Proc. of 2nd Workshop on Roles and Relationships in Object Oriented Programming, Multiagent Systems, and Ontologies. pp. 39-46 (2007)
9. Mascardi, V., Cordì, V., Rosso, P.: A Comparison of Upper Ontologies. In: Web-Oriented Architectures, pp. 55-64. (2007)
10. Masolo, M., Borgo, S., Gangemini, A., Guarino, N., Oltramari, A., Oltramari, A.: WonderWeb Deliverable D18 – Ontology Library (final). Technical report, ISTC-CNR (2004)
11. Mommers, L.: Ontologies in the Legal Domain. In: Theory and Applications of Ontology: Philosophical Perspectives, pp. 264-276. Springer, Berlin Heidelberg (2010)
12. Puleston, C., Parsia, B., Cunningham, J., Rector, A.: Integrating Object-Oriented and Ontological Representations: A Case Study in Java and OWL, ISWC (2008)
13. Robertson, B: Exploring Historical RDF with HEML. Digital Humanities Quarterly, 3(1). (2009). http://www.digitalhumanitites.org/dhq /vol/003/1/000026/000026.html
14. Sartor, G., Casanovas, P., Biasiotti, M., Fernández-Barrera. M.: Approaches to Legal Ontologies: Theories, Domains, Methodologies, Springer (2010).